\documentclass{article}
\usepackage{graphicx} 
\usepackage{natbib}
\usepackage{a4wide}

\newcommand{\apj}{The Astrophysical Journal}

\newcommand{\aap}{Astronomy \& Astrophysics}

\newcommand{\mnras}{Monthly Notices of the Royal Astronomical Society}
\newcommand{\nat}{Nature}

\newcommand{\icarus}{Icarus}

\title{Comment on ``Did the terrestrial planets of the Solar System form by
pebble accretion?''}
\author{Anders Johansen\\ {\small Globe Institute, University of Copenhagen} \\Peter Olson\\ {\small Earth and Planetary Sciences, University of New Mexico} \\Zachary Sharp\\ {\small Earth and Planetary Sciences, University of New Mexico}}
\date{November 2024}

\begin{document}

\maketitle

\section{Abstract}

Morbidelli, Kleine \& Nimmo (2024) (MKN) recently published a critical analysis
on whether the terrestrial planets in the Solar System formed by rapid pebble
accretion or by the classical route of multiple giant impacts between planetary
embryos after the dissipation of the protoplanetary disc. They arrive at the
conclusion that the terrestrial planets  did not form by pebble accretion.
Although we welcome debate on this topic, we want to emphasize here several
points that we disagree on. We will not address in detail every claim made in
MKN, but rather stick to four main points. Our conclusion is that pebble
accretion remains a viable mechanism to drive significant growth of protoplanets
in the protoplanetary disc, with as much as 70\% of Earth formed by pebble
accretion. This rapid growth phase must nevertheless have been followed by an
extended period of collisional growth after the end of the protoplanetary disc
phase, likely culminating with the moon-forming giant impact. We emphasize here
an important recent result from Olson \& Sharp (2023), namely that significant
growth by pebble accretion can be reconciled with the Hf-W decay system even for
a canonical moon-forming giant impact with a Mars-mass protoplanet and a low
equilibration efficiency -- a more massive impactor, as proposed in Johansen
et al.\ (2023), is not necessary. Given that terrestrial planet formation
naturally involves both pebble accretion and a combination of small and large
impactors, this challenges the very notion of making an either/or distinction
between the classical collision model and the pebble accretion model. 

\section{Conditions for pebble accretion}

A circumstellar disk rotates at a sub-Keplerian velocity around the central star to compensate for the outward pressure exerted by the gas. Small particles are coupled with the gas but are not affected by the outward gas pressure -- they will therefore spiral towards the central star. Pebble accretion refers to the process whereby a seed mass orbiting around the star accretes this infalling material leading to a rapid growth of a planetesimal and planetary embryo (Ormel \& Klahr, 2010; Lambrechts \& Johansen, 2012). For this process to be relevant to the early growth of our terrestrial planets, three conditions must be met: 1) a seed mass must be generated early in the protoplanetary disc history; 2) the protoplanetary disc must persist long enough for the planetesimal to grow; and 3) there must be sufficient pebble-size material to grow the planets. The top end of the initial mass function of planetesimals formed by the streaming instability will contain some bodies large enough to grow by pebble accretion (Liu et al., 2020; Lyra et al., 2023). The solar protoplanetary disc persisted for at least 3 Myr based on chondrule ages (e.g., Villeneuve et al., 2009), which is sufficient time to grow an Earth-sized body. And finally, the density and sizes of the solids in the protoplanetary disc structures are sufficient to grow terrestrial planets. Therefore, pebble accretion is predicted to operate from basic physical principles.

\section{The isotopic dichotomy}

MKN discuss in detail how the isotopic composition of Earth lies on a mixing line between meteorites from the inner Solar System (commonly dubbed the NC group) and meteorites from the outer Solar System (dubbed the CC group) for some major lithophile elements such as Cr, Ti, Ca and Si. This is commonly used as an argument in favour of pebble accretion (Schiller et al., 2018), since inward drift of pebbles provides a fundamental physical mechanism that could have transported material from the outer Solar System into the inner Solar System.

MKN then go on to review that for other elements, such as Mo and Zr, Earth does not lie on a mixing line between NC and CC. They briefly discuss our model from Onyett et al. (2023) where it was proposed that Earth and Mars were enriched in presolar SiC grains rich in s-process isotopes. We proposed this to be the result of thermal processing of FeS in the envelope of a growing planet and release of certain elements embedded in the FeS, leading to escape of a small fraction (5\%-10\%) of elements such as Mo and Ru hosted in the FeS component. This process will naturally enrich the growing planet in s-process-enhanced isotopes hosted in the more refractory SiC grains. We maintain here that the model from Onyett et al. (2023) explains well why Earth does  not lie on a mixing line between the NC and CC components in elements such as Mo. Alternatively, one has to evoke that Earth formed out of a class of planetesimals that are not sampled in the current meteorite collections (Burkhardt et al., 2021). We find that the need to invoke such a massive yet hidden ``lost reservoir'' to explain Earth's peculiar Mo composition constitutes a significant problem for a classical collision model without any pebble contribution.

MKN also put forward Zn as an example of an element whose isotopic composition does not agree with pebble accretion, since it appears that Earth's Zn mainly sampled the inner Solar System reservoir  (Kleine et al., 2023). However, Zn is a moderately volatile element with a sublimation temperature of around 700 K. In Steinmeyer et al.\ (2023) we showed that thermal processing of FeS to release H$_2$S, at similar temperatures to Zn sublimation, explains the low S content of Earth. A similar process likely occurred for Zn. Accretion of Zn was quenched after proto-Earth's envelope reached the sublimation temperature of Zn at approximately 0.1 $M_{\rm E}$ -- and the recondensed nanoparticles would escape easily back to the protoplanetary disc with the convective gas plumes (Popovas et al., 2019). Hence it is natural that Earth's depleted Zn reservoir was accreted during the earliest growth phases predominantly from an NC isotopic reservoir. This early accreted Zn was protected from escape from the planetary magma ocean because of the condensation of silicate clouds over the magma ocean that led to the emergence of an impenetrable radiative zone (see Steinmeyer \& Johansen, 2024, who based their 1-D envelope model on the convection criterion of Guillot, 1995 and Leconte et al., 2017).

Further details on this and other related topics on the connection between cosmochemistry and planet formation are given in our forthcoming review in Nature Reviews Chemistry on ``The cosmochemistry of planetary systems'' (Bizzarro et al., 2025).

\section{The Hf-W system}

MKN present a detailed analysis of the Hf-W system following rapid pebble accretion and a later moon-forming giant impact (Yu \& Jacobsen, 2011). They criticize the result of Johansen et al.\ (2023) where we showed that an impactor of mass 0.4 $M_{\rm E}$ with a metal-silicate equilibration efficiency of 25\%-100\% explains our planet's low excess of $^{182}$W relative to $^{183}$W. This is because for such giant impacts, much of the impactor's metallic core may simply merge with the proto-Earth's core, resulting in very low equilibration with the mantle of proto-Earth (Landeau et al., 2021). We largely agree with MKN that a high equilibration fraction and a large impactor is needed for a pure pebble accretion model supplemented by a single moon-forming impact to agree with the Hf-W system of Earth. However, MKN did not properly discuss the implications of an important recent result by Olson \& Sharp (2023). In that model, the Earth grows to 70\% of its final mass during the protoplanetary disc phase and subsequently accretes 20\% of its mass from small impactors (with 90\% equilibration between these small impactor's metal cores and the magma) and finally experiences a giant impact with a Mars-mass body using a conservative 10\% equilibration efficiency. This gives a good match to the $^{182}$W excess of Earth. These two models are illustrated in Figure 1.

We agree with MKN that the low excess of $^{182}$W in the mantle of Earth is evidence of significant growth after the dissipation of the protoplanetary disc. However, this does not imply that pebble accretion is insignificant. In fact, as we discussed above, hybrid pebble accretion / collisional models demonstrate that proto-Earth may have grown to as much as 70\% of its final mass within the protoplanetary disc (Olson \& Sharp, 2023). Importantly, we emphasize that it not a necessity in the pebble accretion model to have a near-equal-mass moon-forming giant impact.

\begin{figure}[!t]
    \centering
    \includegraphics[width=\linewidth]{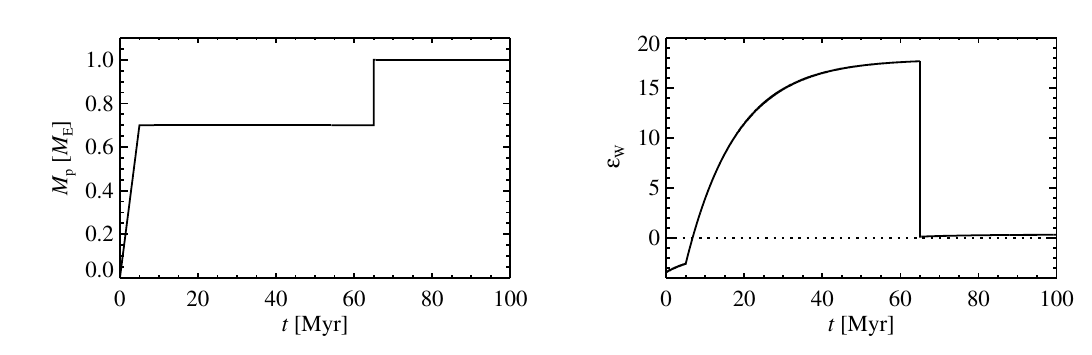}
    \includegraphics[width=\linewidth]{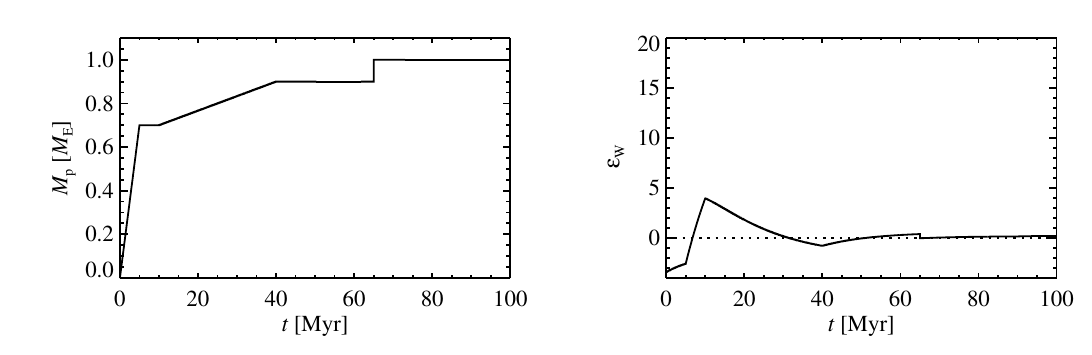}
    \caption{The top panel shows an Earth growth model similar to Johansen et al.\ (2023). Earth here grows to 0.7 $M_{\rm E}$ in the protoplanetary disc and experiences a moon-forming giant impact after 65 Myr. The excess of $^{182}$W (denoted $\epsilon_{\rm W}$) is lowered to the terrestrial value (zero by definition) after the giant impact. We nevertheless assumed here complete equilibration between impactor core and magma ocean, which may not be realistic. In the bottom panel we reproduce the model of Olson \& Sharp (2023). Here Earth grows again to 0.7 $M_{\rm E}$ by pebble accretion in the protoplanetary disc and then accretes an additional 0.2 $M_{\rm E}$ remnant planetesimals between 10 Myr and 40 Myr; this lowers the $^{182}$W excess significantly. The giant impact with a Mars-mass body is here able to lower the $\epsilon_{\rm W}$ value down the terrestrial level even assuming a conservative equilibration efficiency between the impactor core and the silicate magma of only 10\%.}
    \label{fig:enter-label}
\end{figure}

\section{Planetesimals in the Solar System}

MKN thoroughly discuss whether the isotopic compositions of meteorite parent bodies in the Solar System are consistent with a temporal evolution of the transition from CC to NC, or whether these two components were completely isolated from each other, perhaps by the gap formed by Jupiter. Understanding the evolution of the solar protoplanetary disc and the timing of planetesimal formation is clearly a daunting task, and a variety of interpretations are possible. For an update of the view we put forward in Liu et al.\ (2022), namely that the composition of the inner Solar System changed with time from NC to CC due to pebble drift, we refer the reader to Colmenares et al.\ (2024), where we show that a temporal evolution of the composition of the protoplanetary disc following an FU Orionis outburst agrees well with the formation of planetesimals in the end member NC and CC components as well as in intermediate compositions.

\section{Earth's elemental composition}

MKN also claim that pebble accretion cannot explain Earth's depletion in some moderately volatile elements, and that this depletion is better explained by the formation of planetesimals under elevated temperatures in the inner regions of the protoplanetary disc (Sossi et al., 2022). However, more recently, Steinmeyer et al.\ (2023) showed that thermal processing during pebble accretion can deplete a growing planet in moderately volatile elements. In the case of S, Steinmeyer et al.\ (2023) showed that the reaction FeS + H$_2$ = Fe + H$_2$S releases the ultra-volatile molecule H$_2$S, which easily escapes from the envelope by diffusion. We used a similar argument in Johansen et al.\ (2021) to explain the loss of C- and N-bearing compounds from the envelope after their release from accreted organic molecules, as well as in Onyett al.\ (2023) where we used thermal processing to explain the loss of FeS-hosted Mo from the envelope. Although none of these individual studies fully explores the depletion pattern of Earth's moderately volatile elements, they define a promising path for future exploration of explaining the volatile depletion pattern of Earth as a consequence of pebble sublimation during the accretion.

An important difference between pebble accretion and the traditional model of impacts between planetary embryos is that pebble accretion selectively incorporates mm-size pebbles (chondrules, refractory inclusions, metal grains) but mostly excludes dust (matrix in chondrites). As a result, in Garai et al.\ (2024) we are able to duplicate the Earth’s major element composition using pebble components, but cannot do the same with any combination of chondrites (C-chondrite, E-chondrite, O-chondrite and iron meteorite mixing). This is because the pebble accretion model is expected to incorporate early-formed refractory inclusions, while the chondrite-based model, because chondrites formed relatively late, mostly excludes such objects.

\section{Conclusion}

Our overall view is that a hybrid model -- in which the terrestrial planets formed by a combination of pebble accretion, collisions with planetesimals, and giant impacts -- can explain most of the properties of the terrestrial planets in the Solar System. Astrophysical, cosmochemical, and geochemical evidence on the formation of the terrestrial planets is nevertheless ambiguous by nature, so that multiple interpretations are always possible. The actual division of planetary growth between pebble accretion and collisional accretion will depend on the temperature and density conditions in the protoplanetary disc, the life-time of the disc, the strength of the turbulence, the presence of one or several giant planets that may each partially block the pebble flux, the sizes of the pebbles, as well as the efficiencies of planetesimal and embryo formation. Many outcomes are possible, depending on the possible ranges of these properties. Accordingly, in our opinion the best way to make progress is to investigate the interplay among these factors. Planetesimal accretion and giant impacts were always key growth processes even in pebble accretion models  (see Levison et al., 2015; Johansen et al., 2021; Onyett al., 2023; Olson \& Sharp, 2023). We thus do not find it fruitful to overemphasize a strong distinction in rocky planet formation theory between EITHER pebble models OR collision models. Reality is most likely somewhere in between in the terrestrial planet zone where both pebble accretion rates and planetesimal/protoplanet impact rates are high.

\end{document}